\documentclass[prb,showpacs,twocolumn,preprintnumbers,amsmath,amssymb,floatfix]{revtex4}

\usepackage{graphicx,color}

\begin{document}

\title{Diffusion quantum Monte Carlo calculation of the quasiparticle
  effective mass of the two-dimensional homogeneous electron gas}

\author{N.\ D.\ Drummond}

\affiliation{Department of Physics, Lancaster University, Lancaster LA1 4YB,
  United Kingdom}

\author{R.\ J.\ Needs}

\affiliation{TCM Group, Cavendish Laboratory, University of Cambridge,
  J.\ J.\ Thomson Avenue, Cambridge CB3 0HE, United Kingdom}

\date{\today}

\begin{abstract}
The quasiparticle effective mass is a key quantity in the physics of electron
gases, describing the renormalization of the electron mass due to
electron-electron interactions.  Two-dimensional electron gases are of
fundamental importance in semiconductor physics, and there have been numerous
experimental and theoretical attempts to determine the quasiparticle effective
mass in these systems. In this work we report quantum Monte Carlo results for
the quasiparticle effective mass of a two-dimensional homogeneous electron
gas.  Our calculations differ from previous quantum Monte Carlo work in that
much smaller statistical error bars have been achieved, allowing for an
improved treatment of finite-size effects.  In some cases we have also been
able to use larger system sizes than previous calculations.
\end{abstract}


\maketitle


\section{Introduction}

Two-dimensional (2D) electron gases are ubiquitous in modern semiconductor
devices. Surprisingly, however, there remain significant gaps in our knowledge
of the properties exhibited by these fascinating systems.  In recent years
experimentalists have realized increasingly high-quality (low-disorder) 2D
homogeneous electron gases (HEGs) at low densities in quantum-well
structures\cite{padmanabhan,gokmen} and field-effect
transistors.\cite{tan_2005} As the density is lowered, correlation effects
play an ever more important role, and it is hoped that the resulting exotic
behavior could be exploited in a new generation of electronic and spintronic
devices.  At very low densities the Coulomb repulsion between the electrons
dominates, and the HEG forms a Wigner
crystal.\cite{wigner,tanatar,rapisarda,ndd_2D_wigner} At higher densities the
kinetic energy dominates and the electrons form a Fermi fluid, in which most
properties are qualitatively (and at high densities, quantitatively) similar
to those of a free-electron gas.  \textit{Fermi liquid
  theory}\cite{landau,giuliani} is the phenomenological framework within which
the normal behavior of Fermi fluids is understood.

According to Fermi liquid theory, low-lying excitation energies in HEGs are
free-electron-like, and the effects of interactions are encapsulated in (i) a
renormalization of the electron mass (the quasiparticle effective mass) and
(ii) a set of parameters describing the interaction of pairs of excited
quasiparticles. In this work we use quantum Monte Carlo (QMC) methods to
determine the quasiparticle effective mass by calculating single-particle
excitation energies as differences in the total energy when electrons are
either added to or removed from the ground state.

In our calculations we have used the variational Monte Carlo (VMC) and
diffusion Monte Carlo (DMC) methods.\cite{foulkes_2001} In VMC we take the
expectation value of the many-electron Hamiltonian with respect to a
Slater-Jastrow-backflow trial wave
function,\cite{ndd_jastrow,kwon_bf,backflow} which is optimized by minimizing
first the variance of the energy,\cite{umrigar_1988a,ndd_newopt} then the
energy expectation value\cite{umrigar_emin} with respect to free parameters in
the wave function. In DMC\cite{ceperley_1980} we simulate a population of
``walkers'' whose dynamics are governed by the Schr\"{o}dinger equation in
imaginary time in order to project out the ground-state component of an
initial wave function.  The fixed-node approximation\cite{anderson_1976} is
used to impose fermionic antisymmetry.  All our QMC calculations were
performed using the \textsc{casino} code.\cite{casino_ref}

In Ref.\ \onlinecite{ndd_band} we presented a DMC calculation of the 2D HEG
single-particle energy band, enabling us to predict the quasiparticle
effective mass.  In the present work we have had access to the Jaguar machine
at Oak Ridge Leadership Computing Facility, enabling us to achieve higher
accuracy in our DMC calculations, and leading to a refinement of our earlier
work.

The rest of this paper is structured as follows.  In
Sec.\ \ref{sec:landau_e_fnal} we give an overview of the relevant aspects of
Fermi liquid theory.  In Sec.\ \ref{sec:qmc_calcs} we describe our
computational approach.  Our results are presented in Sec.\ \ref{sec:results}.
Finally, we draw our conclusions in Sec.\ \ref{sec:conclusions}. We use
Hartree atomic units, in which the Dirac constant, the electronic charge and
mass, and $4\pi$ times the permittivity of free space are unity
($\hbar=|e|=m_e=4\pi\epsilon_0=1$), throughout.

\section{Landau energy functional \label{sec:landau_e_fnal}}

\subsection{Parameterization of excitation energies}

According to Fermi liquid theory the total energy of a HEG in a particular
excited state is given by the Landau energy functional\cite{giuliani}
\begin{eqnarray} E & = & E_0+\sum_{{\bf k},\sigma} {\cal E}_\sigma({\bf k})
  \delta {\cal N}_{{\bf k},\sigma} \nonumber \\ & & {} + \frac{1}{2}
  \sum_{({\bf k},\sigma) \neq ({\bf k}^\prime,\sigma^\prime)}
  f_{\sigma,\sigma^\prime}({\bf k},{\bf k}^\prime) \delta {\cal N}_{{\bf
      k},\sigma} \delta {\cal N}_{{\bf
      k}^\prime,\sigma^\prime}, \label{eq:landau_e_fnal}
  \end{eqnarray}
where $\delta {\cal N}_{{\bf k},\sigma}$ is the change to the ground-state
quasiparticle occupation number for wavevector ${\bf k}$ and spin $\sigma$,
and $E_0$ is the ground-state energy.  The energy band ${\cal E}_\sigma({\bf
  k})$ is the energy of an isolated quasiparticle.  Near the Fermi surface,
the energy band may be assumed to be linear and hence we may write
\begin{equation} {\cal E}_\sigma({\bf k}) = {\cal
  E}_F+\frac{k_F}{m^\ast}(k-k_F), \label{eq:qem_defn} \end{equation} where
${\cal E}_F$ is the Fermi energy, $k_F$ is the Fermi wavevector, and $m^\ast$
is the quasiparticle effective mass.  The Landau interaction function
$f_{\sigma,\sigma^\prime}({\bf k},{\bf k}^\prime)$ describes energy
contributions arising from pairs of quasiparticles, and will not be considered
further in this paper.

The goal of this work is to obtain accurate values for the 2D HEG
quasiparticle effective mass $m^\ast$ in the thermodynamic limit at different
densities and for different spin polarizations, giving us the most important
contribution to the Landau energy functional.

\subsection{Spin-polarization effects}

Both theoretical work\cite{zhang_2005,ndd_band} and experimental
studies\cite{padmanabhan,gokmen} have shown that the quasiparticle effective
mass has a significant dependence on the spin polarization of the HEG\@.  We
have calculated the effective mass for both paramagnetic and ferromagnetic
(fully spin-polarized) HEGs.  Fully spin-polarized HEGs are experimentally
relevant because they may be created by applying an in-plane magnetic field to
a 2D electron system.  Differences in the quasiparticle effective masses of
ferromagnetic and paramagnetic HEGs result in differences in the transport
properties, which could be exploited in electronic or spintronic applications,
e.g., in devices that use the spin-Coulomb-drag effect.\cite{scd}

\subsection{Finite-size errors}

The 2D HEGs encountered in real devices are sufficiently large that they can
be regarded as being of essentially infinite extent.  In QMC simulations we
can only study small numbers of electrons, however.  For a HEG in a finite
simulation cell subject to periodic boundary conditions, momentum quantization
limits the available wavevectors $\{{\bf k}\}$ to a discrete lattice.
Furthermore, long-range Coulomb and correlation effects cannot be treated
exactly in a finite cell,\cite{holzmann_2009,holzmann_2011} giving rise to
finite-size errors in the energy band and hence effective mass.

Fermi liquid theory is only valid for excitations near the Fermi surface: in
this region the quasiparticle lifetime becomes large and hence the
quasiparticle momentum occupancies are good quantum numbers.\cite{giuliani}
The energy band is defined by the Landau energy functional at all ${\bf k}$,
but does not correspond to the quasiparticle band except in the vicinity of
the Fermi surface.  In the infinite-system limit, the exact energy band is
smooth in general and, if the quasiparticle effective mass is well-defined,
the band must be at least differentiable at the Fermi surface. 

\section{QMC calculations \label{sec:qmc_calcs}}

\subsection{Choice of simulation cell}

In all our calculations the simulation cell was square and the simulation-cell
Bloch vector\cite{rajagopal_1994,rajagopal_1995} was ${\bf k}_s={\bf 0}$.  The
number of electrons in the ground state was chosen to give a closed-shell
configuration in each case.  For ferromagnetic HEGs, our calculations were
performed with $N=29$, $57$, and $101$ electrons in the ground state.  For
paramagnetic HEGs our calculations were performed with $N=26$, $50$, $74$, and
$114$ electrons in the ground state.

The simulation cell was identical for all excitations of a given HEG; hence
the electron density increased when electrons were added and decreased when
electrons were removed from the ground-state configuration.  This procedure
results in zero finite-size error for a free-electron gas.

\subsection{Trial wave functions}

We use real, single-determinant trial wave functions for the closed-shell
ground states, which is a computationally efficient approach that facilitates
the optimization of the wave function.  In our QMC calculations we used
Slater-Jastrow-backflow trial wave functions. The Jastrow factors consisted of
polynomial and plane-wave expansions in the interelectron
distances,\cite{ndd_jastrow} while the backflow functions consisted of
polynomial expansions in the interelectron distances.\cite{backflow} The
polynomial expansions were cut off smoothly at the radius of the largest
circle that could be inscribed in the simulation cell. The Jastrow factor and
backflow function contained a total of 35 and 17 free parameters,
respectively, for paramagnetic HEGs, and 27 and 8 free parameters,
respectively, for ferromagnetic HEGs.  Extrapolation of the VMC energy with
different trial wave functions to zero energy variance\cite{ndd_band} suggests
that our DMC calculations retrieved more than 99\% of the correlation energy.

For each density, system size, and spin polarization the wave function was
optimized in the ground state and the resulting Jastrow factor and backflow
function were used in all the excited states, with the exception of a couple
of test cases, as discussed in Sec.\ \ref{sec:wf_reopt}.

\subsection{DMC time steps, etc.}

The DMC time steps used in our calculations were $0.04$, $0.2$, and $0.4$
a.u.\ at $r_s=1$, 5, and 10, respectively, for paramagnetic HEGs, and $0.01$,
$0.2$, and $0.4$ a.u.\ at $r_s=1$, 5, and 10, respectively, for ferromagnetic
HEGs.  It was verified that halving the time step had a negligible effect on
the energy band: leading-order time-step errors cancel out of the total-energy
differences involved.  The target population exceeded 1200 configurations in
each case, ensuring that population-control bias is negligible.

The number of equilibration steps discarded from the start of each DMC
calculation was sufficiently large that the root-mean-square distance diffused
by each electron in the equilibration period exceeded the linear size of the
simulation cell.

\section{Results \label{sec:results}}

\subsection{Energy band}

\subsubsection{DMC results for the energy band \label{sec:eband_DMC_results}}

\begin{figure}
\begin{center}
\includegraphics[clip,width=0.4\textwidth]{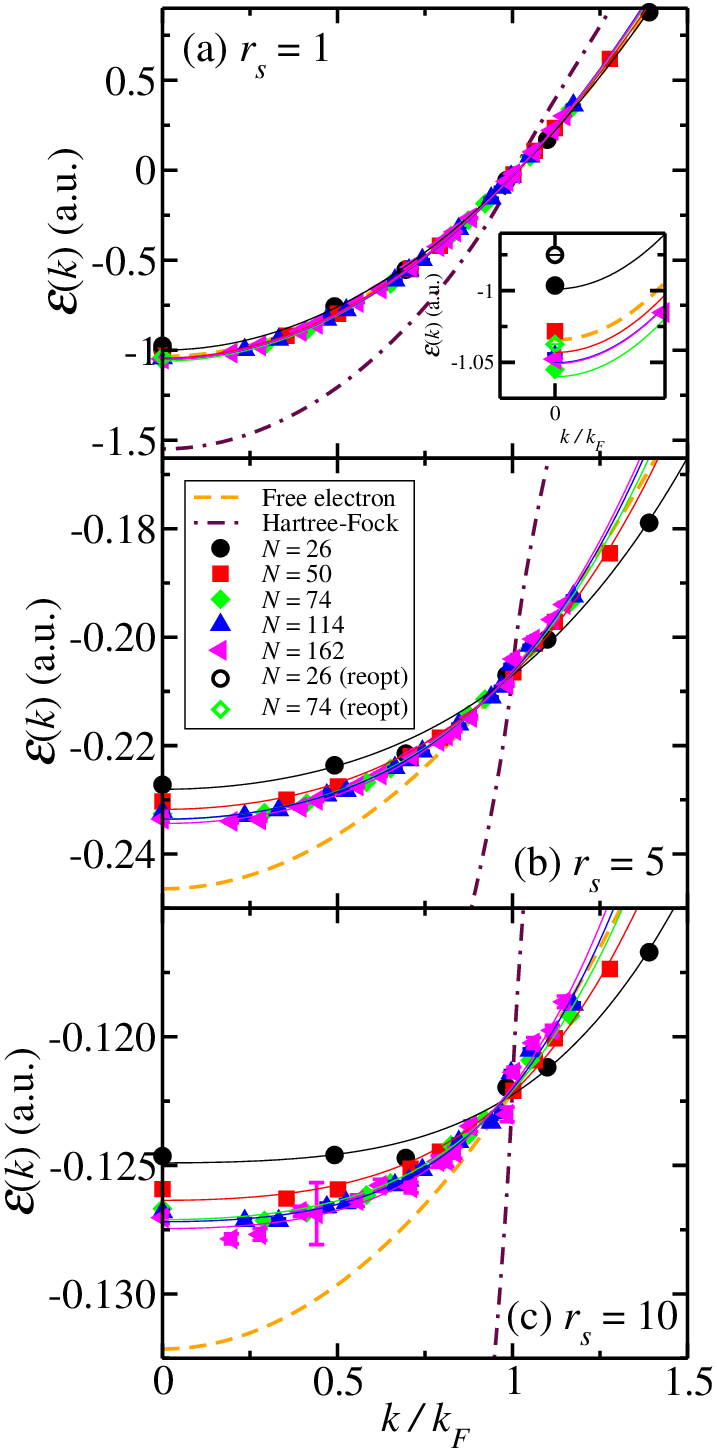}
\caption{(Color online) Energy bands ${\cal E}({\bf k})$ for paramagnetic 2D
  HEGs of density parameter (a) $r_s=1$, (b) $r_s=5$, and (c) $r_s=10$ at
  different system sizes $N$. For the curves labeled ``reopt,'' the wave
  function was optimized separately in the ground state and excited states.
  The free-electron and Hartree-Fock bands have been offset so that they
  coincide with the DMC bands at $k_F$. The inset to panel (a) shows the
  energy band around ${\bf k}={\bf 0}$ in greater
  detail. \label{fig:para_energy_bands}}
\end{center}
\end{figure}

\begin{figure}
\begin{center}
\includegraphics[clip,width=0.4\textwidth]{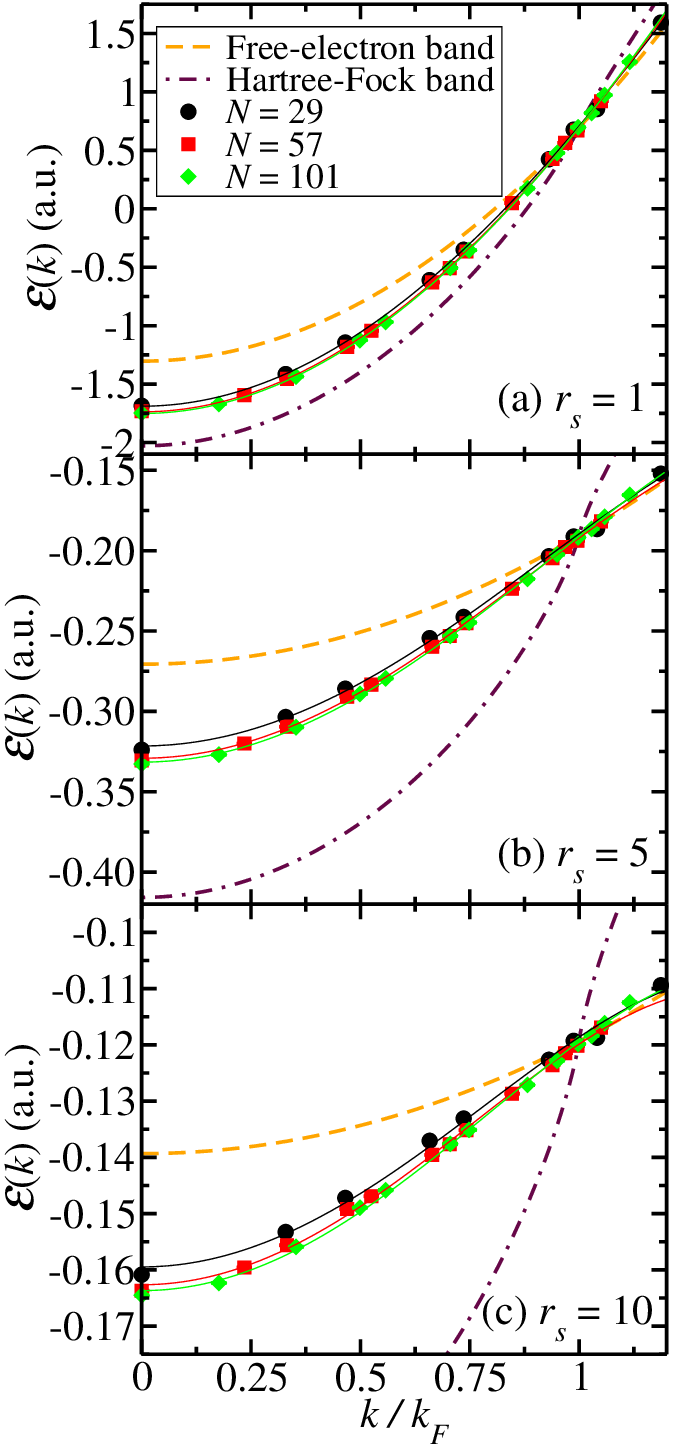}
\caption{(Color online) As Fig.\ \ref{fig:para_energy_bands}, but for
  ferromagnetic HEGs. \label{fig:ferro_energy_bands}}
\end{center}
\end{figure}

The DMC energy band of the paramagnetic 2D HEG, obtained with a
Slater-Jastrow-backflow trial wave function, is shown in
Fig.\ \ref{fig:para_energy_bands}.  Analogous results for a fully
ferromagnetic HEG are shown in Fig.\ \ref{fig:ferro_energy_bands}.

The energy bands calculated in this work at any given system size are in
agreement with those calculated in Ref.\ \onlinecite{ndd_band}, but the
statistical error bars in the present work are very much smaller due to the
considerably larger computational resource available.  Furthermore, the random
noise in the trial wave function due to optimization by energy minimization
with a finite sampling of configuration space is greatly reduced because of
the enormous numbers of configurations that were used in the optimizations.
As a result it is now possible to discern a systematic trend in the energy
band with system size, with the bandwidth tending to increase with system size
$N$.  This in turn leads to a reduction in the predicted quasiparticle
effective mass in the thermodynamic limit, as discussed in
Sec.\ \ref{sec:qem}.

\subsubsection{Effect of reoptimizing the wave function in excited
  states \label{sec:wf_reopt}}

The excitation of a single electron or pair of electrons has no effect on the
optimal Jastrow factor or backflow function in the thermodynamic limit; hence
the fact that the Jastrow factor and backflow function can be reoptimized in
an excited state in a finite cell is simply a manifestation of finite-size
error. Reoptimizing the wave function when an electron was subtracted from
${\bf k}={\bf 0}$ in a 26-electron HEG at $r_s=1$ lowered the DMC energy by
$0.000854(4)$ a.u., reducing the DMC bandwidth and hence increasing the
finite-size error, as can be seen for $N=26$ in the inset of
Fig.\ \ref{fig:para_energy_bands}(a). In our calculations we therefore
optimized the trial wave function in the ground state and then continued to
use the same Jastrow factor and backflow function in our excited-state
calculations.

\subsection{Quasiparticle effective mass \label{sec:qem}}

\subsubsection{Quartic fits to the energy bands}

At each density, spin polarization, and system size a quartic function ${\cal
  E}(k)=\alpha_0+\alpha_2 k^2+\alpha_4 k^4$ was fitted to the DMC energy-band
values.  The effective mass [defined in Eq.\ (\ref{eq:qem_defn})] was then
calculated as
\begin{equation} m^\ast=\frac{k_F}{(d{\cal
      E}/dk)_{k_F}}. \label{eq:qem_eval} \end{equation} 

\begin{figure}
\begin{center}
\includegraphics[clip,width=0.4\textwidth]{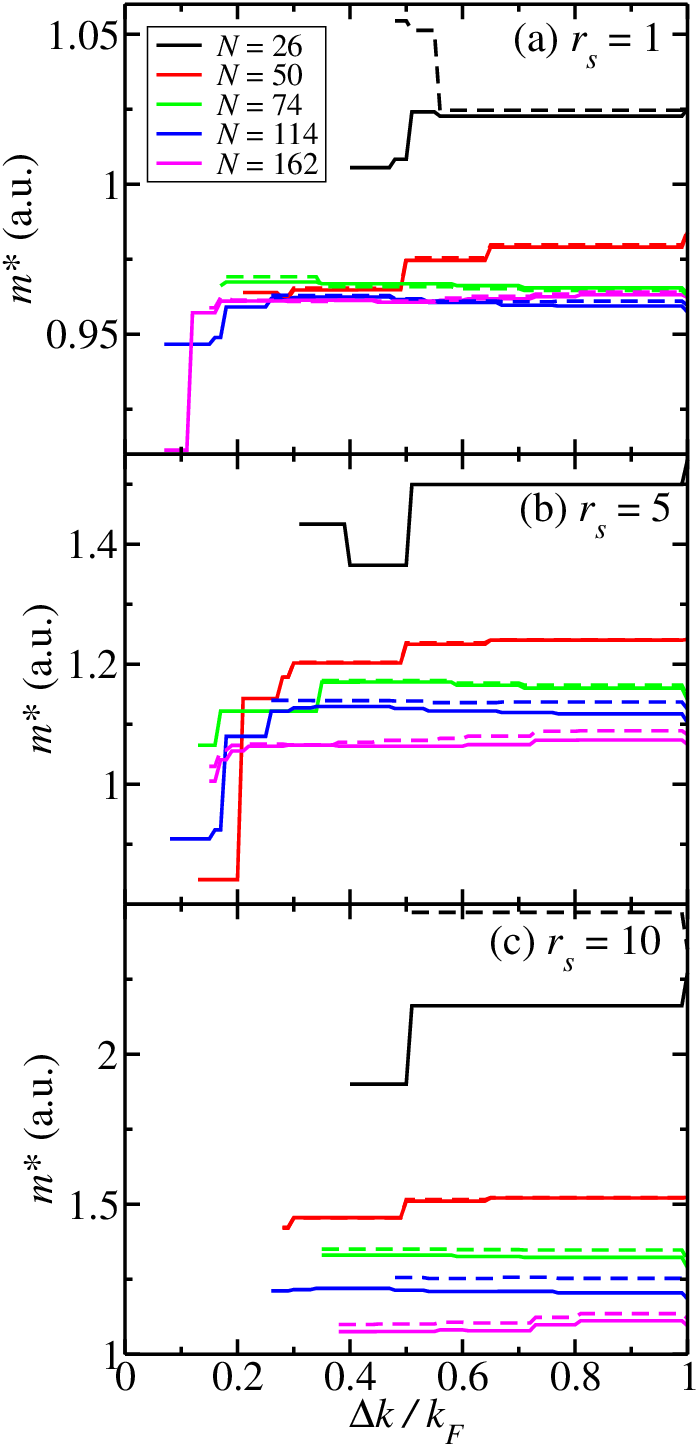}
\caption{(Color online) Quasiparticle effective mass $m^\ast$ against range of
  wavevectors included in the quartic fit of the energy band. Specifically,
  only wavevectors in the interval $[k_F-\Delta k,k_F+\Delta k]$ are used in
  the fit.  For the dashed lines, wavevectors within 10\% of $k_F$ are
  excluded from the fit.
\label{fig:para_qem_v_deltak}}
\end{center}
\end{figure}

\begin{figure}
\begin{center}
\includegraphics[clip,width=0.4\textwidth]{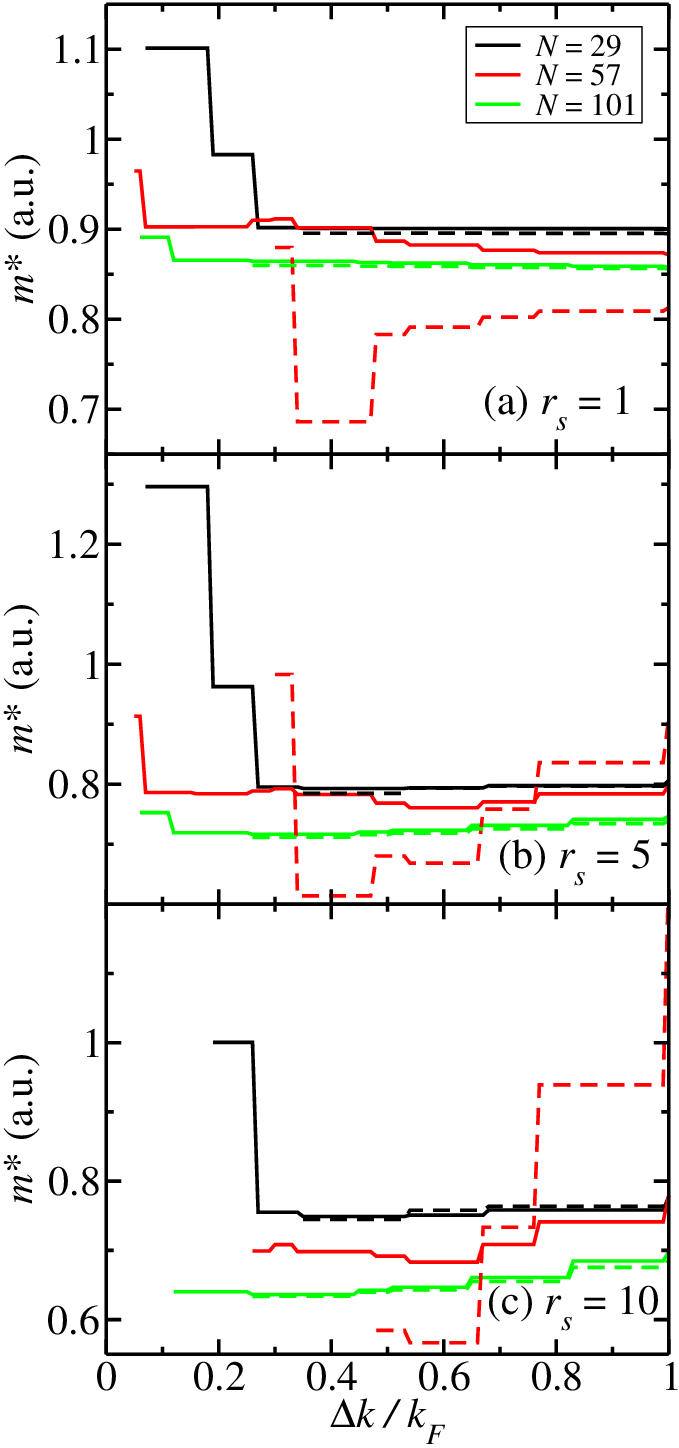}
\caption{(Color online) As Fig.\ \ref{fig:para_qem_v_deltak}, but for
  ferromagnetic HEGs.  For $N=57$ electrons, excluding wavevectors within 10\%
  of $k_F$ eliminates all the wavevectors above $k_F$ at which energy-band
  data are available.
\label{fig:ferro_qem_v_deltak}}
\end{center}
\end{figure}

We have investigated the dependence of the estimate of the effective mass on
the range of energy-band data used to perform the fit.  Figures
\ref{fig:para_qem_v_deltak} and \ref{fig:ferro_qem_v_deltak} show the
effective mass as a function of the range $\Delta k$ about the Fermi wave
vector over which we perform the fit.  The figures also show the effective
mass when energy-band data from within $\pm 10$\% of $k_F$ are excluded from
the fit.  It is clear that the effective mass becomes pathological when
$\Delta k$ becomes small (i.e., only excitations in the vicinity of $k_F$ are
considered), for the reasons discussed briefly in Sec.\ \ref{extrapolation}
and at length in Ref.\ \onlinecite{ndd_band}.  As the width of the region over
which the fit is performed becomes larger, the effective mass estimates settle
down to well-defined values that behave in a systematic fashion with system
size.  There is no evidence of any need to exclude data from around $k_F$
however: effectively, the fitting process averages out the pathological
behavior in the vicinity of the Fermi surface.

\subsubsection{Extrapolation to the thermodynamic limit \label{extrapolation}}

\begin{figure}
\begin{center}
\includegraphics[clip,width=0.45\textwidth]{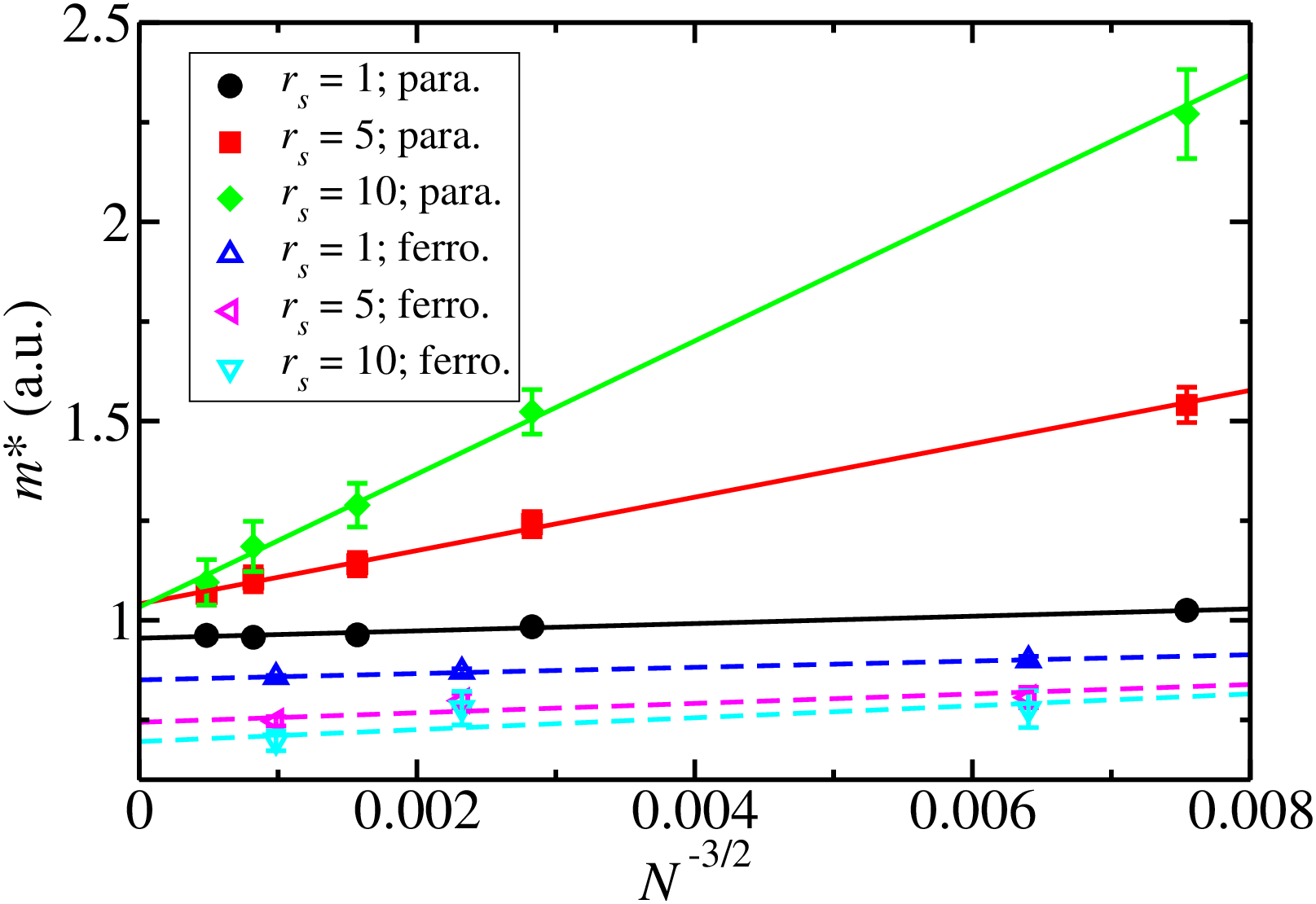}
\caption{(Color online) Quasiparticle effective mass $m^\ast$ against
  $N^{-3/2}$, where $N$ is the system size, for 2D
  HEGs. \label{fig:qem_scaling}}
\end{center}
\end{figure}

The quasiparticle effective masses are plotted against system size in
Fig.\ \ref{fig:qem_scaling}.  A systematic trend in the effective mass as a
function of system size can be seen.  Hence we are able to extrapolate the
effective mass to the thermodynamic limit, significantly reducing finite-size
errors.  The effective-mass values that we report in this work are expected to
be more accurate than those reported in Ref.\ \onlinecite{ndd_band}.

The finite-size error does not show the $N^{-1/4}$ behavior predicted by
Holzmann \textit{et al.}\cite{holzmann_2009}\ for excitations near the Fermi
surface, presumably because we have fitted to the entire band.  (Any
$N^{-1/4}$ behavior in the band near the Fermi surface is masked by the
pathological behavior that is seen in Hartree-Fock theory and hence
QMC\@.\cite{ndd_band}) To investigate the behavior of the effective mass as a
function of system size, we have fitted the function
\begin{equation}
  m^\ast(N)=m^\ast(\infty)+bN^{-\gamma}, \label{eq:qem_extrap} \end{equation}
where $m^\ast(\infty)$, $b$, and $\gamma$ are fitting parameters, to the raw
data shown in Fig.\ \ref{fig:para_qem} and we have performed repeated fits to
the data with Monte Carlo sampling of the error bars.  We find that
$\gamma=1.8(4)$, $1.4(4)$, and $1.4(3)$ at $r_s=1$, $5$, and $10$,
respectively.  This indicates that the optimal exponent $\gamma$ is between 1
and 2.  This conclusion is reinforced by the results shown in Table
\ref{table:fitting_analysis}, where we examine the $\chi^2$ values of the fits
and the extrapolated effective masses when different exponents are used in
Eq.\ (\ref{eq:qem_extrap}).  The extrapolation shown in
Fig.\ \ref{fig:qem_scaling} assumes an exponent of $\gamma=3/2$.

\begin{table}
\begin{center}
\caption{Extrapolated quasiparticle effective mass $m^\ast(\infty)$ and
  $\chi^2$ value for fits to the effective-mass data as a function of system
  size for paramagnetic HEGs with different exponents $\gamma$ in the
  finite-size fitting formula
  [Eq.\ (\ref{eq:qem_extrap})]. \label{table:fitting_analysis}}
\begin{tabular}{cr@{.}lr@{.}lr@{.}lr@{.}lr@{.}lr@{.}l}
\hline \hline

 & \multicolumn{6}{c}{$m^\ast(\infty)$ (a.u.)} & \multicolumn{6}{c}{$\chi^2$
  (a.u.)} \\

\raisebox{1.5ex}[0pt]{$\gamma$} & \multicolumn{2}{c}{$r_s=1$} &
\multicolumn{2}{c}{$r_s=5$} & \multicolumn{2}{c}{$r_s=10$} &
\multicolumn{2}{c}{$r_s=1$} & \multicolumn{2}{c}{$r_s=5$} &
\multicolumn{2}{c}{$r_s=10$} \\

\hline

$1/4$ & ~$0$&$88$ & ~~$0$&$3$ & ~$-0$&$7$ & ~$21$&$5$ & ~~$9$&$6$ & ~~$12$&$4$
\\

$1/2$ & $0$&$924$ & $0$&$74$ & $0$&$3$  & $16$&$8$ & $5$&$8$  & $7$&$7$  \\

$1$   & $0$&$947$ & $0$&$97$ & $0$&$85$ & $9$&$1$  & $1$&$11$ & $1$&$52$ \\

$3/2$ & $0$&$955$ & $1$&$04$ & $1$&$03$ & $5$&$2$  & $0$&$28$ & $0$&$32$ \\

$2$   & $0$&$959$ & $1$&$08$ & $1$&$13$ & $5$&$3$  & $2$&$03$ & $3$&$1$  \\

\hline \hline
\end{tabular}
\end{center}
\end{table}

\subsubsection{Quasiparticle effective mass as a function of density}

\begin{figure}
\begin{center}
\includegraphics[scale=0.4,clip]{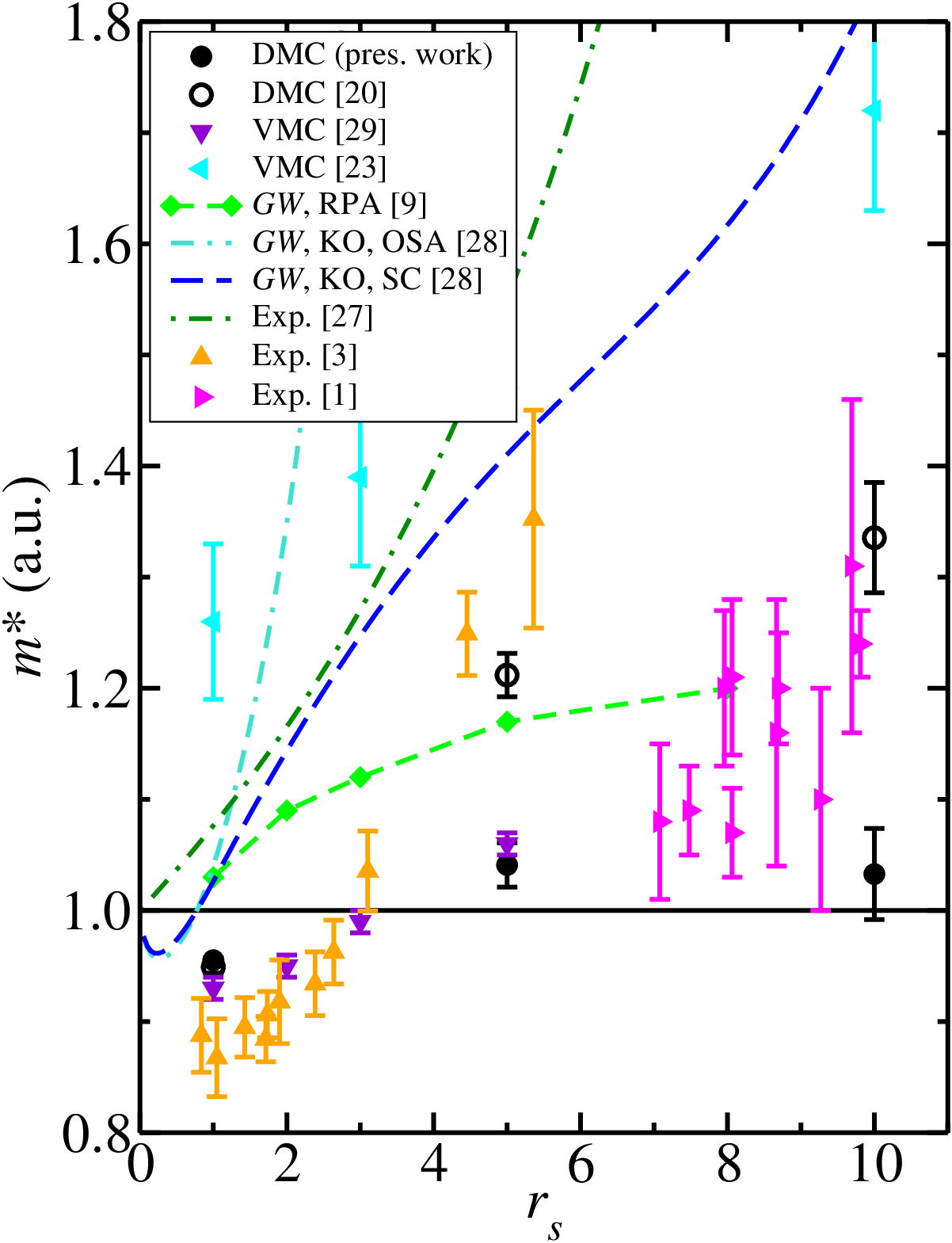}
\caption{(Color online) Quasiparticle effective mass $m^\ast$ against density
  parameter $r_s$ for paramagnetic or partially spin-polarized 2D HEGs, as
  calculated or measured by different authors.  The experimental results are
  due to Smith and Stiles\cite{smith}, Tan \textit{et al.},\cite{tan_2005} and
  Padmanabhan \textit{et al.}\cite{padmanabhan} The $GW$ results were obtained
  using the random-phase-approximation (RPA) effective
  interaction\cite{giuliani} and the Kukkonen-Overhauser (KO) effective
  interaction\cite{asgari_2006} by solving the Dyson equation
  self-consistently (SC) or within the on-shell approximation (OSA)\@.  We
  show the VMC results of Kwon \textit{et al.}\cite{kwon_1994}\ (which were
  later confirmed at the same system size at $r_s=1$ a.u.\ using
  transient-estimate DMC calculations\cite{kwon_trans_est}), the VMC results
  of Holzmann \textit{et al.},\cite{holzmann_2009} and the DMC results
  reported in our previous work,\cite{ndd_band} as well as the results of the
  present work. All the results shown are for paramagnetic HEGs with the
  exception of the experimental results of Ref.\ \onlinecite{padmanabhan},
  which are for a partially spin-polarized HEG\@.  \label{fig:para_qem}}
\end{center}
\end{figure}

\begin{figure}
\begin{center}
\includegraphics[scale=0.3,clip]{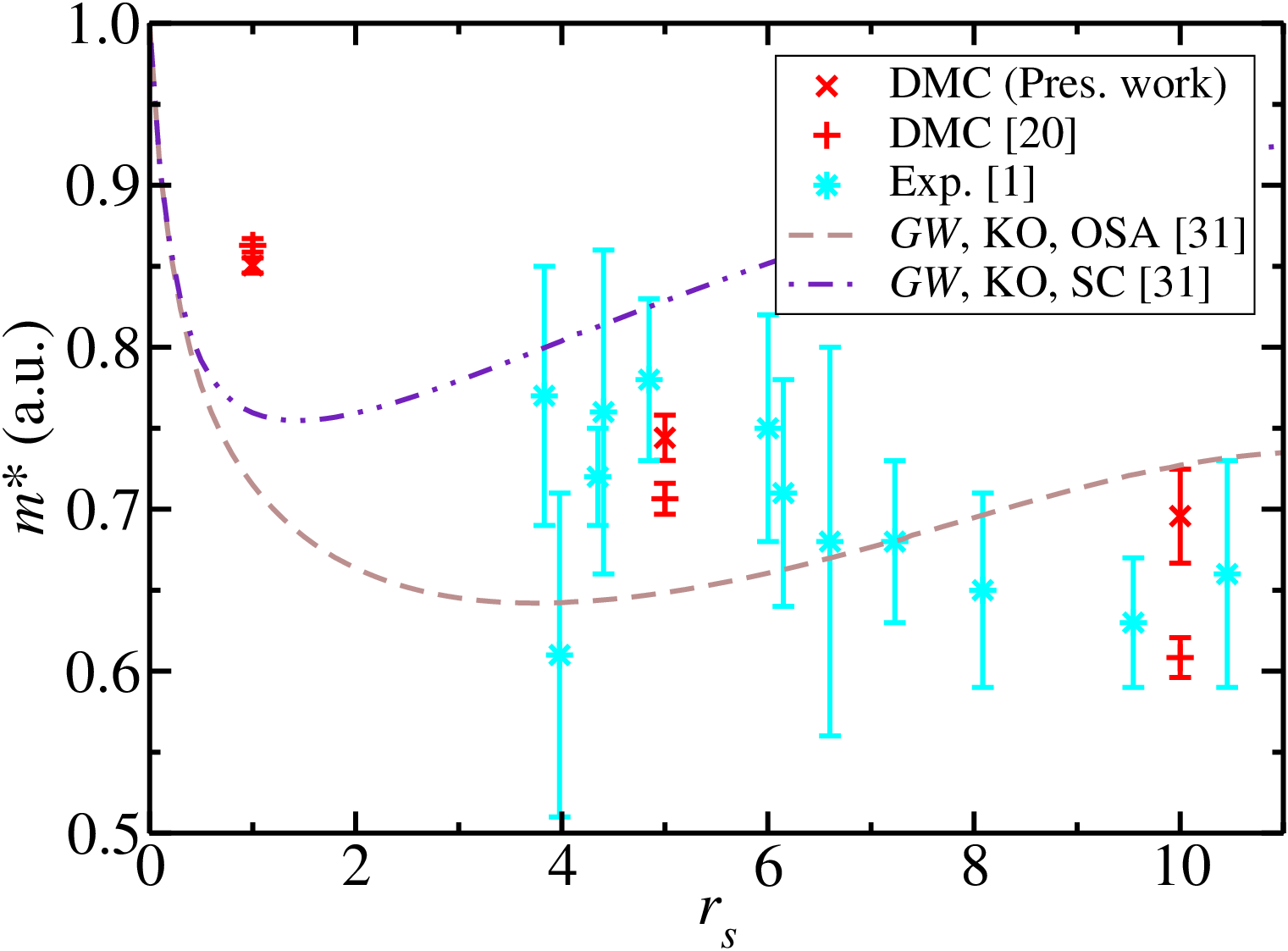}
\caption{(Color online) Quasiparticle effective mass $m^\ast$ against density
  parameter $r_s$ for ferromagnetic 2D HEGs. The $GW$ results were obtained
  using the Kukkonen-Overhauser (KO) effective interaction by solving the
  Dyson equation self-consistently (SC) or within the on-shell approximation
  (OSA)\@.\cite{asgari_2009} The experimental results are due to Padmanabhan
  \textit{et al.}\cite{padmanabhan} We show the DMC results reported in our
  earlier work\cite{ndd_band} in addition to our current results.
\label{fig:ferro_qem}}
\end{center}
\end{figure}

Results for the 2D HEG quasiparticle effective mass obtained by different
authors, including the present work, are shown in Figs.\ \ref{fig:para_qem}
and \ref{fig:ferro_qem} for paramagnetic and fully ferromagnetic HEGs,
respectively.  The present results are also given in Table
\ref{table:qem_results}. For ferromagnetic HEGs our revised effective masses
are in reasonable agreement with our previously published
results.\cite{ndd_band} For paramagnetic HEGs, however, the finite-size errors
are relatively large at low density, and hence finite-size extrapolation
reduces the effective masses at $r_s=5$ and $10$ by a significant amount.  For
paramagnetic HEGs we now find that the effective mass remains close to 1
(i.e., electron-electron interactions result in almost no renormalization of
the electron mass) at all the $r_s$ values we have considered.

\begin{table}
\begin{center}
\caption{Quasiparticle effective masses for paramagnetic and fully
  ferromagnetic 2D HEGs, extrapolated to the thermodynamic
  limit. \label{table:qem_results}}
\begin{tabular}{ccr@{.}l}
\hline \hline

Mag.\ state & $r_s$ & \multicolumn{2}{c}{$m^\ast$ (a.u.)} \\

\hline

Para.  & ~$1$ & $0$&$955(2)$ \\

Para.  & ~$5$ & $1$&$04(2)$  \\

Para.  & $10$ & $1$&$03(4)$  \\

Ferro. & ~$1$ & $0$&$851(5)$ \\

Ferro. & ~$5$ & $0$&$74(1)$  \\

Ferro. & $10$ & $0$&$70(3)$  \\

\hline \hline
\end{tabular}
\end{center}
\end{table}

Once again we emphasize that there is no significant disagreement between the
\textit{data} reported in the present article and Ref.\ \onlinecite{ndd_band}.
The revision of the effective mass simply results from the fact that the
random noise in our current data is much smaller, allowing a systematic trend
with system size to be discerned and hence removed by extrapolation.

For the paramagnetic HEG, GW calculations\cite{giuliani,asgari_2006} indicate
a steep increase in the effective mass as the density is lowered, similar to
that seen in early experiments.\cite{smith} However, the GW results depend
strongly on the choice of effective interaction and whether or not the
calculations are performed self-consistently.  The QMC calculations of
Holzmann \textit{et al.}\cite{holzmann_2009} give quite different results from
those of either Kwon \textit{et al.},\cite{kwon_1994} our previous
work,\cite{ndd_band} or the present work.  The experimental
data\cite{tan_2005,padmanabhan} show some evidence for enhancement of the
effective mass at low density, although we do not see this in our present
results.

For the ferromagnetic case, our effective-mass data are in agreement with the
experimental results of Padmanabhan \textit{et al.},\cite{padmanabhan} showing
a decrease in the effective mass as the density is lowered.  GW
theory\cite{asgari_2009} also predicts a suppression of the effective mass in
the range of densities considered.  However, the difference between the GW
results obtained self-consistently and in the on-shell approximation is
significant, as is the difference with the present results.

\section{Conclusions \label{sec:conclusions}}

We have used DMC to calculate the single-particle energy band and hence
quasiparticle effective mass of the 2D HEG\@.  We have achieved sufficiently
high precision in our calculations that systematic finite-size errors in the
quasiparticle effective mass can be observed and removed by extrapolation.
This leads to a revision of the effective masses for paramagnetic HEGs at low
density compared to our earlier work:\cite{ndd_band} in particular we find
that there is no enhancement of the effective mass at low density.

\begin{acknowledgments}
Financial support was received from Lancaster University under the Early
Career Small Grant Scheme, and the Engineering and Physical Sciences Research
Council.  This research used resources of the Oak Ridge Leadership Computing
Facility at the Oak Ridge National Laboratory, which is supported by the
Office of Science of the U.S.\ Department of Energy under Contract
No.\ DE-AC05-00OR22725.  Additional computing resources were provided by the
Cambridge High Performance Computing Service.
\end{acknowledgments}

\end{document}